\documentclass[twocolumn,prb,floatfix,showpacs,superscriptaddress]{revtex4-1} %

\usepackage{color,amsmath, amssymb, graphicx, amsfonts ,  setspace}

\newcommand{\bra}[1]{\ensuremath{\left\langle #1\right|}}
\newcommand{\ket}[1]{\ensuremath{\left|#1\right\rangle}}

\newcommand{\abs}[1]{\ensuremath{\left| #1 \right|}}

\newcommand{\nn}{\nonumber}
\newcommand{\mi}{\ensuremath{\mathrm{i}}}

\begin{document}

\title{Localisation, delocalisation, and topological transitions in
  disordered 2D quantum walks} 

\author{Jonathan M. Edge}
\affiliation{Nordita, KTH Royal Institute of Technology and Stockholm
  University, Roslagstullsbacken 23 106 91 Stockholm, Sweden}
\author{Janos K. Asboth} \affiliation{ Institute for Solid State
  Physics and Optics, Wigner Research Centre, Hungarian Academy of
  Sciences, H-1525 Budapest P.O. Box 49, Hungary}

\date{\today}
\begin{abstract}
We investigate time-independent disorder on several two-dimensional
discrete-time quantum walks.  We find numerically that, contrary to
claims in the literature, random onsite phase disorder, spin-dependent
or otherwise, cannot localise the Hadamard quantum walk; rather, it
induces diffusive spreading of the walker. In contrast, split-step
quantum walks are generically localised by phase disorder. We explain
this difference by showing that the Hadamard walk is a special case of the
split-step quantum walk, with parameters tuned to a critical
point at a topological phase transition.
We show that the topological phase transition can also be reached by introducing strong disorder in the rotation angles.
We determine the critical
exponent for the divergence of the localisation length at the topological phase transition,
and find $\nu=2.6$, in both cases.  This places the two-dimensional
split-step quantum walk in the universality class of the quantum Hall
effect.
\end{abstract}

\pacs{73.43.Nq, 05.40.Fb, 03.67.Ac}
\maketitle

Discrete-time quantum walks\cite{kempe_2003}, which we will simply
refer to as quantum walks, are the quantum analogues of classical
random walks. They are model systems which sit at the interface
between quantum information theory and condensed matter physics. On
the one hand, they form archetypical systems for studying quantum
algorithms\cite{PhysRevA.67.052307, PhysRevA.79.012325}. On the
other hand, condensed matter physics has recently also shown interest
in quantum walks \cite{Kitagawa2010a,Obuse2011a,Asboth2012} in
particular ever since it was shown that the topological phases
\cite{Hasan2010a,Qi2011} can also be realised in quantum walks
\cite{Kitagawa2010}.

Condensed matter physics has a very wide scope, but one important
subject of it is disorder and the associated localisation of
single-particle wave functions (for a review, see
Ref.~\onlinecite{Evers2008}). Thus, to understand quantum walks from
the condensed matter point of view, we need to address the effect of
disorder on the propagation of a quantum walker.

One of the interesting aspects of quantum walks is that in the absence
of disorder the quantum walker propagates ballistically
\cite{kempe_2003}, thus much faster than its classical counterpart,
which shows diffusive propagation. The ballistic spreading of the
quantum walk is related to the quantum speed up of certain quantum
algorithms, notably Grover's search algorithm\cite{Grover1997}, as the
quantum walker is able to explore the search space more rapidly than
its classical counterpart.

If disorder is introduced into the quantum walk system, it is expected
to break the ballistic propagation of the quantum walk, analogously to
the way in which disorder introduced into a solid-state system will
affect electrons due to disorder
scattering.  This could be of relevance to the quantum information
applications of quantum walks, as the quantum speed up of quantum
algorithms is intimately related to ballistic propagation.

Although the effects of disorder on one-dimensional quantum walks have
been extensively studied, not much is known about the two-dimensional
case. For one-dimensional quantum walks it has been shown that spatial
disorder can lead to exponential localisation of all energy
eigenstates
\cite{Yin2008,Schreiber2011,Ahlbrecht2011,Joye2012,Ghosh2014}.  It was
also found, however, that chiral symmetry can prevent localisation in
one dimension \cite{Obuse2011a}.  To the best of our knowledge, the
effects of spatial disorder in two-dimensional quantum walks and its
impact on the quantum walk propagation was only studied in
Ref.~\onlinecite{Svozilik2012} for the Hadamard walk. In that paper it
was reported that in the disordered system the wave function remains
majoritatively close to the starting position, unlike in the clean
case, where the amplitude of the wavefunction at the initial site
decreases to zero in the long time limit. This concentration of the
wave function close to its initial position (which is according to a
looser terminology used as the definition of localisation, as, e.g, in
Ref.~\onlinecite{Inui2004}) was attributed to Anderson localisation.

In this article we study of the effects of spatial disorder on the
propagation and localisation of the Hadamard walk, and on the broader
family of two-dimensional split-step walks to which it
belongs. Sec.~\ref{sec:defin-quant-walks} collects the definitions of
these walks, recalls their connection and their topological phases. In
Sec.~\ref{sec:phase_disorder} we show that phase disorder localises
generic split-step walks, but not the Hadamard walk: this latter shows
slow diffusion (contrary to the findings of
Ref.~\onlinecite{Svozilik2012}).  In
Sec.~\ref{sec:topological_transition} we attribute this difference to
the fact that the Hadamard walk is critical: it is a split-step walk
that is tuned to a topological phase transition point. We demonstrate
this phase transition and calculate the corresponding critical
exponent, $\nu=2.6$, which places the split-step walk in the quantum Hall
universality class. Finally, in Sec.~\ref{sec:disorder_angle} we
study disorder in the angle parameters of the split-step walks. Based
on the preceding section, one can expect that if the angle disorder is large
enough, the split-step walk can become diffusive even with maximal
phase disorder. We show that this disorder-induced delocalisation
actually takes place, and find for it the same critical exponent $\nu$
as in Sec.~\ref{sec:topological_transition}. We also find that angle
disorder alone leads to diffusion rather than localisation, which is
probably connected to the presence of a particle-hole symmetry 
in this disordered quantum walk.

\section{Definitions of the quantum walks}
\label{sec:defin-quant-walks}

A particle undergoing a quantum walk on a square lattice is
represented by a time-dependent two-component wavefunction, 
\begin{align} 
\ket{\psi(t)} &= \sum_{m,n} \sum_{s=\pm1}
\psi(t)_{m,n,s}\ket{m,n,s}.
\end{align} 
Here $m,n \in \mathbb{Z}$ give the horizontal and vertical positions
on the lattice, $s\in \{+1,-1\}$ is the value of the internal state
that we call spin, and $t\in\mathbb{N}$ denotes the time, which is
only allowed to take on discrete values. We take as initial condition
a localized state, $\ket{0,0,+1} $, and obtain the time evolution by
iterated applications of the time evolution operator $U$ on the state,
\begin{align}
\ket{\psi(t)}&= U^{t} \ket{0,0,+1}. 
\end{align}
We will consider different types of quantum walks, with the time
evolution operator $U$ consisting of a product of several shift
operators and coin operators, to be defined below.

Shift operators displace the walker by one lattice site in a direction
that depends on its internal state, but their action is independent of
the position of the walker.
We consider the quantum walk on a square lattice with the sites
labelled by $(m,n) $ and so define the following shift operators,
\begin{align}
  \hat S_x &= \sum_{m,n}\sum_{s=\pm 1} \ket{m+s,n,s}\bra{m,n,s}; \nn\\ 
  \hat S_y &= \sum_{m,n}\sum_{s=\pm 1}
  \ket{m,n+s,s}\bra{m,n,s}. \nn
\end{align}
We use absorbing boundary conditions\cite{asboth_edge_2014a} in both the
$x$ and $y$ directions.

Coin operators act locally on the walker, but can have
position-dependent parameters. They can be written in compact form
using the Pauli operators, $\sigma_z
\ket{m,n,s}=s\ket{m,n,s}$; $\sigma_x \ket{m,n,s}=\ket{m,n,-s}$;
$\sigma_y \ket{m,n,s}=is\ket{m,n,-s}$; and $\sigma_0
\ket{m,n,s}=\ket{m,n,s}$, for all values of $m,n$ and $s$.  We
consider the Hadamard coin operator 
\begin{align}
\hat H &= 2^{-1/2} \big( \sigma_x + \sigma_z \big).
\end{align}
and the spin rotation
operator, 
\begin{align}
\hat R[\theta_j]&=
  \sum_{m,n} e^{-\mi \theta_j^{mn} \sigma_y }\ket{m,n}\bra{m,n},  \nn
\end{align}
with $\theta^{mn}_j$ denoting the position-dependent rotation
angles. The index $j$ differentiates between rotations in one sequence
of operations defining the timestep; below, the time evolution
operator will contain two spin rotations, and so $j$ will take values
1 and 2.  Since $\hat{R}[\theta_j + \pi] = - \hat{R}[\theta_j]$, only
angles between $-\frac\pi2$ and $\frac\pi2$ give distinct rotation
operators (the minus sign is only a phase factor).

The first type of quantum walk we consider is the Hadamard walk, 
defined through its time evolution operator,
\begin{align}
  U_{H} =\hat {S}_y \hat{H}
  \hat{S}_x  \hat{H}.
  \label{eq:hadamard_evol_op}
\end{align}
It thus consists of a Hadamard coin operation followed by a
spin-dependent displacement in the $x$ direction, another Hadamard
coin operation rotation and a displacement in the $y$ direction.

\begin{figure}[tb]
  \centering
  \def \picwidth {\columnwidth}
  \includegraphics[width=\picwidth]{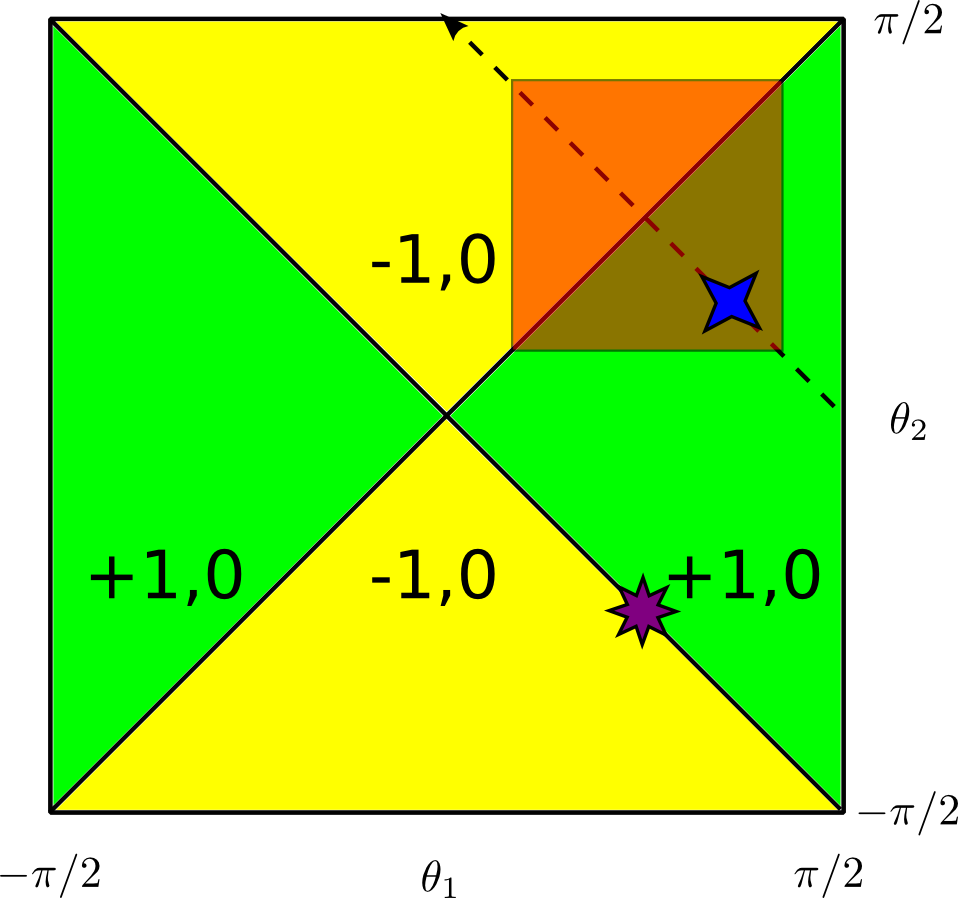}  
  \caption{(Color online) Phase diagram for the topological quantum numbers for the
    split-step quantum walk defined by
    eq.~\eqref{eq:evol_op_splitstep_walk} without disorder
    \cite{asboth_edge_2014a}. As described in
    Ref.~\onlinecite{asboth_edge_2014a}, due to the time-periodic
    nature of the quantum walk, two topological invariants can be
    defined, only one of which changes in the parameter range under
    consideration in this paper. The red transparent box shows the
    range of $\theta_1$ and $\theta_2$ which is accessible at the
    point $\theta_1=\theta_2=\pi/4$ for the parameters in
    fig.~\ref{fig:diffn_coeff_fn_of_eta-whole_range} and
    \ref{fig:distribution_of_nus}.  The blue four-sided star shows the
    parameter set $\theta_1=0.35\pi, \theta_2=0.15\pi$ which is
    frequently used throughout this paper.  The purple eight-sided
    star shows the parameters $\theta_1=\pi/4, \theta_2=-\pi/4$ at
    which the split-step quantum walk reduces to the Hadamard walk.  }
  \label{fig:top_phase_diag}
\end{figure}

We also consider the split-step quantum walk\cite{Kitagawa2010}, where
the time evolution operator is defined as
\begin{align}
  \label{eq:evol_op_splitstep_walk}
  U_s=\hat S_y\hat R[\theta_2] \hat S_x\hat R[\theta_1].
\end{align}
As described in Ref.~\onlinecite{asboth_edge_2014a}, for rotation
angles $\theta_j^{mn}=\theta_j$ independent of position, the system
has two topological invariants: the Chern number and the quasienergy
winding\cite{Rudner2012}, which are determined by $\theta_1$ and
$\theta_2$. The phase diagram for the topological invariants is
reproduced in Fig.~\ref{fig:top_phase_diag}. The Chern number for this
quantum walk is always zero, but, as we will see, the quasienergy winding
plays an important role in determining the localisation properties.

The split-step quantum walk can be seen as a generalisation of the
Hadamard walk. Since $H = \sigma_x e^{-i(\pi/4)\sigma_y} =
e^{i(\pi/4)\sigma_y} \sigma_x$, we have
\begin{align}
U_H &= S_y R(-\pi/4) S_x^{-1} R(\pi/4).
\label{eq:hadamard_special_splitstep}
\end{align}
Thus the Hadamard walk is the a mirror reflected $x \leftrightarrow
-x$ version of the split-step walk, with $\theta_1=-\theta_2=\pi/4$.

\section{The effect of phase disorder}
\label{sec:phase_disorder}

One way to introduce discrete-time independent disorder into quantum walks, is
to multiply the wavefunction at the end of each timestep by a random
phase factor, which depends on position and spin value, but not on
time. For this, we define the phase operators
\begin{align}
  \hat P_a[\phi] &= \sum_{m,n}e^{\mi\phi_{mn} \sigma_a}\ket{m,n}\bra{m,n}, \nn
\end{align}
with $a=0$ for a spin-independent, and $a=z$ for  a spin-dependent phase
operator.  We take the phases $\phi_{mn}$ to have zero mean value, and
distributed randomly in the interval $[-\delta\phi/2,\delta\phi/2)$.
Intuitively, $\hat P_0$ mimics an on-site energy in a tight
binding lattice model, while $\hat P_z$ can be understood as a
disordered magnetic field. 
As such, these types of disorder favour localisation
in non-interacting two-dimensional lattice systems
\cite{Abrahams1979}.

\subsection{Hadamard walk with phase disorder: Disorder-induced diffusion}
\label{subsec:absence-localisation}
To add phase disorder to the Hadamard quantum walk,
Eq.~\eqref{eq:hadamard_evol_op}, we define the timestep operator as 
\begin{align}
  U_{H,a} =\hat{P}_{a} \hat {S}_y \hat{H}
  \hat{S}_x  \hat{H}.
  \label{eq:hadamard_disorder_evol_op}
\end{align}
For different values of $\delta\phi$ between 0 and $2\pi$, and
different disorder realizations, we initialise the quantum walker at
the centre of a 220 $\times$ 220 lattice 
\footnote{Throughout this article we choose lattice sizes large enough
  for the boundary to have no effect.}, and follow the time evolution
for 1000 time steps.

To detect localisation, we will use two of its signatures.  First, in
the presence of localisation, the wave function in the long time limit
should decrease exponentially as a function of the distance from the
initial site,
\begin{align}
\sum_{s=\pm 1} \abs{\Psi(t\to\infty)_{m,n,s}^2} \propto
e^{-2\sqrt{m^2+n^2}/\xi}.
\label{eq:def_xi}
\end{align}
The localisation length $\xi\in\mathbb{R}$ of a localized wavefunction
should be well defined (at least in the vicinity of the initial site).
Second, in the localised case, the spreading $s(t)$ of the wave
function, defined as 
\begin{align}
s^2(t)&=\sum_{m,n} \sum_{s=\pm 1} (m^2+n^2) \abs{\Psi(t)_{m,n,s}}^2,
\end{align}
should saturate, i.e., $\lim_{t\to\infty} s(t) = \text{const}$.

\begin{figure}[tb]
  \centering 
\def \picwidth {\columnwidth}
  \includegraphics[width=\picwidth]{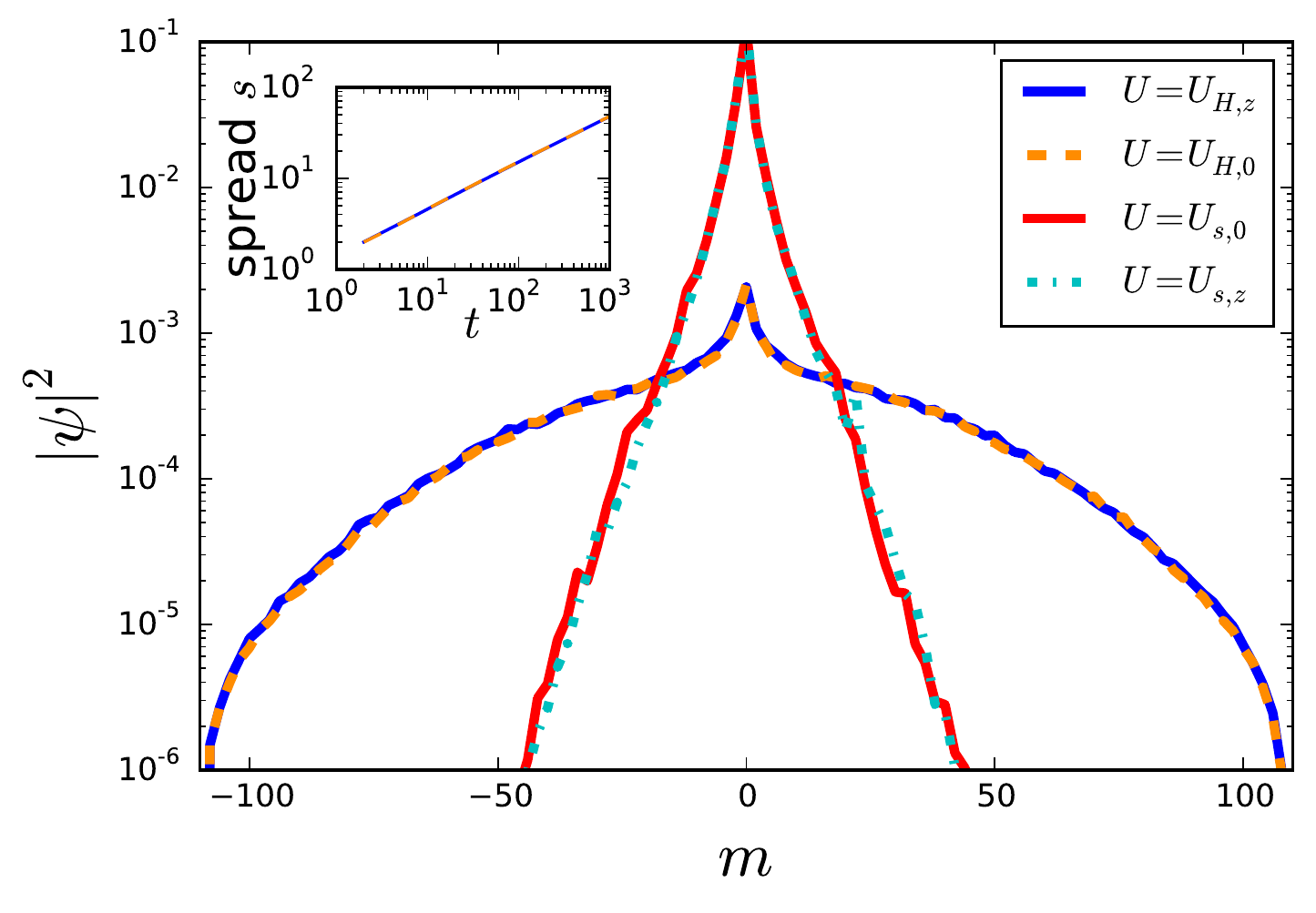}
  \caption{(Color online)  Wavefunction cross-section at $n=0$ after 1000 time steps
    for a quantum walker initialised at the centre of an 220x220
    lattice, averaged over 500 disorder realisations. Only even
    lattice sites are shown, as the wave function is zero on odd
    lattice sites for $t=1000$. In total four types of quantum walks
    are shown: $U_{H,a}$ (Eq.~\eqref{eq:hadamard_disorder_evol_op}) and
    $U_{s,a}$ (Eq.~\eqref{eq:evol_op_with_onsite}) with
    $a=0,z$, $\delta\phi=2\pi$.
    For $U_{s,a}$ we choose $\theta_1=0.35\pi$, $\theta_2=0.15\pi$ and $\delta\theta=0.14\pi$.
    As we can see,
    $U_{H,0}$ and $ U_{H,z}$ show similar types of
    diffusive behaviour.
    The inset shows the spreading of the
    wavefunction for $U_{H,z}$ and $U_{H,0}$ (same color
    coding as in the main figure), which it is roughly consistent with
    the diffusive $s\sim \sqrt t$ behaviour.
    In contrast, $U_{s,0}$ and $U_{s, z}$ show localising behaviour according to Eq.~\eqref{eq:def_xi}.
  }
  \label{fig:wfn_xsection_hadamard}
\end{figure}

In the Hadamard walk with phase disorder, we find diffusive dynamics
instead of localisation.  In Fig.~\ref{fig:wfn_xsection_hadamard} we
have plotted a cross-section of the probability amplitude squared of
the wave function after 1000 timesteps of both the Hadamard walk with
spin-dependent, and spin-independent disorder, averaged over 500
disorder realizations. We see that although the wave function is
strongly peaked towards the centre, it does not decay exponentially:
in both cases, it shows a Gaussian profile characteristic of diffusive
behaviour \cite{Lemarie2009-PRA}. The inset shows the spreading
$s(t)$, which displays no sign of saturation: it is well approximated
by $s(t)\propto t^{1/2}$, which again is an indication of diffusion.

Our results contradict those of Ref.~\onlinecite{Svozilik2012}, where
localisation was found for the disordered Hadamard walk, and also go
against the intuitive picture that onsite disorder induces
localisation. Although it cannot, in principle, be ruled out that
localisation will eventually set in, the 1000 times steps we
considered give an already significantly larger timescale than the 20
time steps investigated in Ref.~\onlinecite{Svozilik2012}. Why is
there no localisation in the disordered Hadamard walk? This is one of
the main questions which we will answer below.

\subsection{Split-step walk with phase disorder: Disorder-induced localisation}
\label{sec:local-due-onsite}

To obtain a full picture of phase disorder and its effects on
localisation, we now apply disorder to the generic split-step walk,
which can be seen as a generalization of the Hadamard walk,
cf.~Eq.~\eqref{eq:hadamard_special_splitstep}. We fix the rotation
angles at $\theta_1=0.35 \pi$ and $\theta_2=0.15 \pi$. As seen on the
phase space of the walk, Fig.~\ref{fig:top_phase_diag}, this set of
parameters is far from the continuous lines along which the
quasienergy gap closes.  The time evolution operator is then given by
\begin{align}
  U_{s,a} = \hat P_a [\phi]  S_y R[\theta_2] S_y R[\theta_1],
  \label{eq:evol_op_with_onsite}
\end{align}
with $a=0$ for spin-independent, and $a=z$ for spin-dependent
disorder. We remark that both types of phase disorder break the
particle-hole symmetry of the system, which arose since $U_s$ was real
\cite{Kitagawa2010}.

\begin{figure}[tb]
  \centering \def \picwidth {\columnwidth}
  \includegraphics[width=\picwidth]{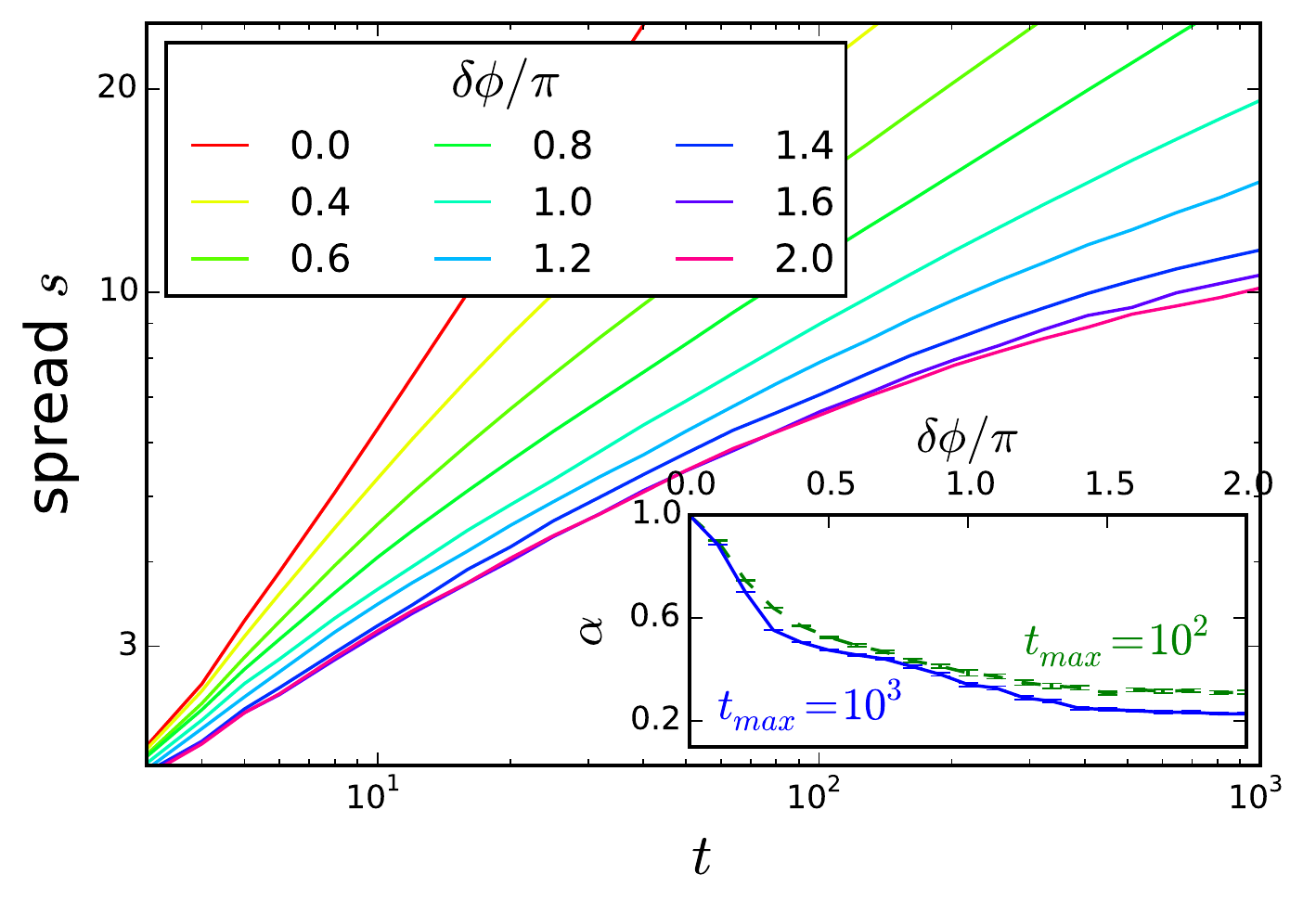}
  \caption{(Color online) Spreading $s(t)$ of the wavefunction in the quantum walk
    $U_{s,0}$ of
    Eq.~\eqref{eq:evol_op_with_onsite} as a function of time, with
    various amounts of phase disorder $\delta \phi$, averaged over 100
    disorder realisations (log-log plot).  The rotation angle
    parameters are set to $\theta_1=0.35\pi, \theta_2=0.15\pi$. Upon
    increasing the phase disorder, the walk shifts from a delocalised
    ($s(t)\propto t^\alpha$) to a localised ($\lim_{t\to\infty} s(t)=
    \text{const}$) behaviour. Inset: exponent $\alpha$ of $s(t)\propto
    t^{\alpha}$ fitted to the curves between $t_{min}=10$ and
    $t_{max}$ (blue solid line: $t_{max}=100$, green dashed line:
    $t_{max}=1000$). For, $\delta\theta=0$, the
    system behaves ballistically, with $\alpha=1$. For larger
    disorder the fitted value of $\alpha$ decreases
    with time, indicating localisation. }
  \label{fig:xover_deloc_loc_as_fn_of_phi}
\end{figure}

Our numerical results indicate that unlike the Hadamard walk, the 2-D
split-step quantum walk is localised by phase disorder.  As shown in
Fig.~\ref{fig:xover_deloc_loc_as_fn_of_phi}, in the absence of phase
disorder, $\delta\phi=0 $, the wave function spreads ballistically, as
expected.  As $\delta\phi $ is increased, however, the wave function
spreads more slowly, and for large values of $\delta\phi$, it seems to
saturate indicating localisation.  The inset of
Fig.~\ref{fig:xover_deloc_loc_as_fn_of_phi} shows the localisation
transition through the exponent $\alpha$ obtained by fitting $s(t)
\propto t^{\alpha}$ to the numerical results over short ($10<t<100$,
green dashed) and long \footnote{For small values of disorder the
  walker spreads very rapidly, such that it reaches the boundary
  before $t=1000$. For these cases we have fitted $\alpha$ in the
  range $10\leq t\leq t_{\delta\phi}$ where $t_{\delta\phi}$ is chosen
  to be a time before the walker has reached the boundary}
($10<t<1000$, blue solid ) times. When $\delta\phi=0$ we observe
ballistic propagation, indicated by a time-independend value of
$\alpha=1$. For increasing values of disorder $\alpha$ decreases and,
more importantly, decreases as a function of time. This indicates that a
power law fit for $s(t)$ does not provide a good fit and that the
system is localising. Additional evidence for localisation is furnished by the shape of the wavefunction in the long-time limit, as shown in Fig.~\ref{fig:wfn_xsection_hadamard}.

\section{Topological transition behind delocalization}
\label{sec:topological_transition}

The difference in the effects of phase disorder on the Hadamard walk
(diffusion) and the generic split-step quantum walk (localisation), is
due to the fact that the Hadamard walk is a special case of the
split-step walk, tuned to a topological phase transition point. In
this Section we expand on this explanation, and investigate it
numerically, obtaining the critical exponents corresponding to this
phase transition via single parameter scaling.

To make sure that the effect we observe is generic, we also include a
small amount of disorder in the angle parameters of the split-step
quantum walk. These angles $\theta_j^{mn}$ will be chosen randomly and
independently for each site, from a uniform distribution in the
interval $[\theta_j-\delta\theta, \theta_j+\delta\theta)$.  Thus the
first and second rotation have the same disorder $\delta\theta$,
which we fix in this section to be $\delta\theta = 0.2\pi$.

\subsection{Topological transition by tuning the mean rotation angles}
\label{sec:top_trans-mean_rot_angles}

We locate the topological phase transition, by tuning the parameters
of the quantum walk: we gradually increase $\theta_2$ from $0$ to
$\pi/2$ while keeping $\theta_1+\theta_2=\pi/2$ constant, all the while
keeping maximal phase disorder, $\delta\phi=2\pi$, and a moderate
angle disorder, $\delta\theta = 0.2\pi$.  This path is marked by the
dashed line in Fig.~\ref{fig:top_phase_diag}.  We characterise
the localisation properties for each set of  parameter values via the
time-dependent diffusion coefficient,
\begin{align}
  D(t)= \frac{s^2(t)}t.
  \label{eq:2}
\end{align}
In the long-time limit the diffusion coefficient $D(t)$ is a constant in regimes governed by diffusion
(metallic or possibly critical regimes) and decreases in time in the
localised regime ('non-metallic' regime). We
choose this quantity because it will be a suitable starting point
for the scaling analysis of the transition point.

\begin{figure}[tb]
  \centering
  \def \picwidth {\columnwidth}
  \includegraphics[width=\picwidth]{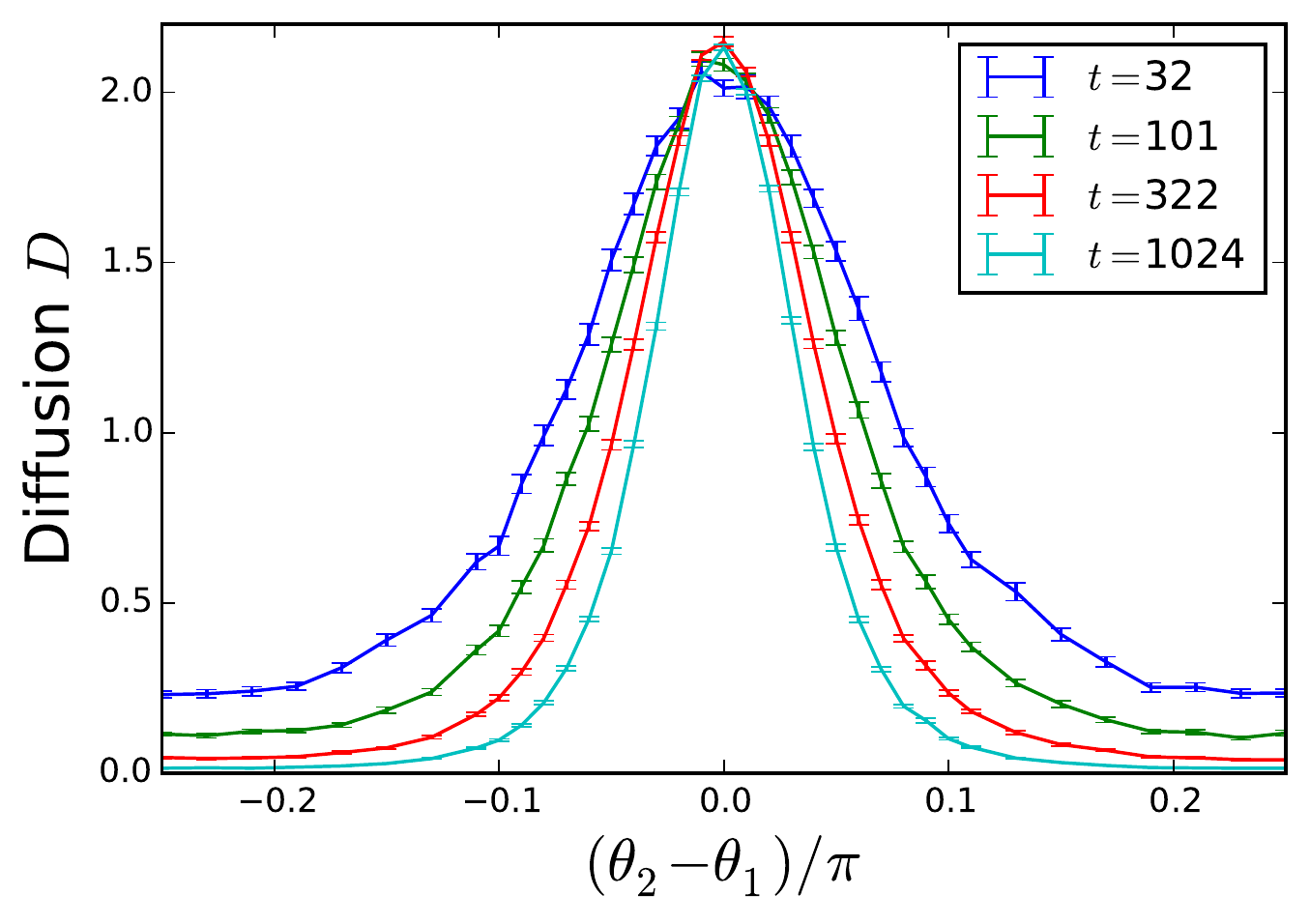} 
  \caption{(Color online) Diffusion coefficient as a function of $\theta_2-\theta_1$ with $\theta_1+\theta_2=\pi/2$ for $ \delta\theta=0.2\pi$, $\delta\phi=2\pi$, 
    obtained by averaging over 100 disorder realisations on a 220x220 lattice.}
  \label{fig:diffn_coeff_fn_of_eta-whole_range}
\end{figure}

Our results for the diffusion coefficient $D$ for various times, as
the rotation angles $\theta_j$ are tuned across topological phase
transition, are shown in
Fig.~\ref{fig:diffn_coeff_fn_of_eta-whole_range}.  At most values of
the angles, the calculated values of $D(t)$ decrease with time $t$,
and we can infer that the quantum walk is localised.  At the point
$\theta_1=\theta_2=\pi/4$, however, the curves of $D (t)$
corresponding to various times overlap, and so the system is
diffusive. This is a delocalisation transition. 

We attribute the delocalisation at $\theta_1=\theta_2=\pi/4$
to the occurrence of a topological phase transition.
In the absence of disorder, the quantum walk has topological invariants $(-1, 0) $,
and $(+1, 0) $  at the endpoints
of the path, respectively. It is thus plausible that somewhere along the path a topological phase transition has to occur. 
Our observations show that this transition occurs at the point $\theta_1=\theta_2=\pi/4$, which is also what one might expect on symmetry grounds.

Another angle from which to understand the delocalisation at
$\theta_1=\theta_2=\pi/4$ is the following.  At the interface between
two domains of the quantum walk with different topological phases there
are edge states \cite{asboth_edge_2014a}.  If both possible topological
quantum numbers occur locally with equal probability, a percolating
network of edge states appears.
At $\theta_1=\theta_2=\pi/4$ the possible local values of $\theta_1$ and $\theta_2$ are shown by the red transparent box in Fig.~\ref{fig:top_phase_diag}. 
This network can be thought of as a
realisation of the Chalker-Coddington network model for the integer
quantum Hall effect \cite{Chalker1988}, tuned to the plateau transition
point. At this point a non-zero conductance appears, which in this
case is signalled by a diffusively spreading wavefunction.

\subsection{Scaling analysis of the localisation-delocalisation transition}
\label{sec:char-local-deloc}

In this section we perform a scaling analysis of the transition at
$\theta_1=\theta_2=\pi/4$, where the localisation length $\xi$ of
Eq.~\eqref{eq:def_xi} has to diverge. We use the same approach as for
the corresponding transition in the quantum anomalous Hall
effect\cite{Dahlhaus2011}: we compute high accuracy data for the
diffusion coefficient $D(t)$, and then we fit this data assuming
power-law divergence of the localization length and single-parameter
scaling. We summarize the main ideas and
the results here and relegate the details to
Appendix~\ref{sec:deta-scal-analys}.

The split-step quantum walk with generic values of $\theta_1$ and
$\theta_2$ has a phase-disorder-dependent localisation length $\xi$,
defined in Eq.~\eqref{eq:def_xi}. This quantity effectively determines
how far the wave function may spread. At a topological transition the
localisation length has to diverge (there is no length scale
associated with diffusive, i.e., metallic propagation). We assume that
this divergence happens as a power law, in analogy with the quantum
Hall case \cite{Slevin2009},
\begin{align}
\xi&=A\abs{\eta}^{-\nu};
\label{eq:def_nu}
\end{align}
Here $\eta$ is the distance from the critical point, $A$ is a constant
of proportionality and $\nu$ is the critical exponent\cite{Evers2008}.
When this transition is obtained as explained above, along the
line $\theta_1+\theta_2=\pi/2$, the role of $\eta$ is played by
\begin{align}
\eta &= \theta_2-\theta_1. 
\label{eq:def_eta}
\end{align}

Instead of measuring the localisation length $\xi$ directly (which would require a calculation of $D(t)$ up to much larger times), we find
$\nu$ by assuming single-parameter scaling of the diffusion
coefficient $D(t)$ of Eq.~\eqref{eq:2}. Taking finite-time
corrections\cite{Lemarie2009-PRA} into account, we have
\begin{align}
  \label{eq:scaling_ansatz_numeric}
  \ln D(t)&=F(t^{1/2\nu} u) + t^{- y}G(t^{1/2\nu} u); \\ 
    u&=\eta + {\cal O}(\eta ^2).
\end{align}
Here the scaling functions $F(z)$, $G(z)$, and $u(z)$, as well as the
exponents $y$ and $\nu$ are to be fitted to the numerical data. The
quality of the fits will provide justification for the single
parameter scaling assumption.

\begin{figure}[tb]
  \def \picwidth {\columnwidth}
  \centering
  \includegraphics[width=\picwidth]{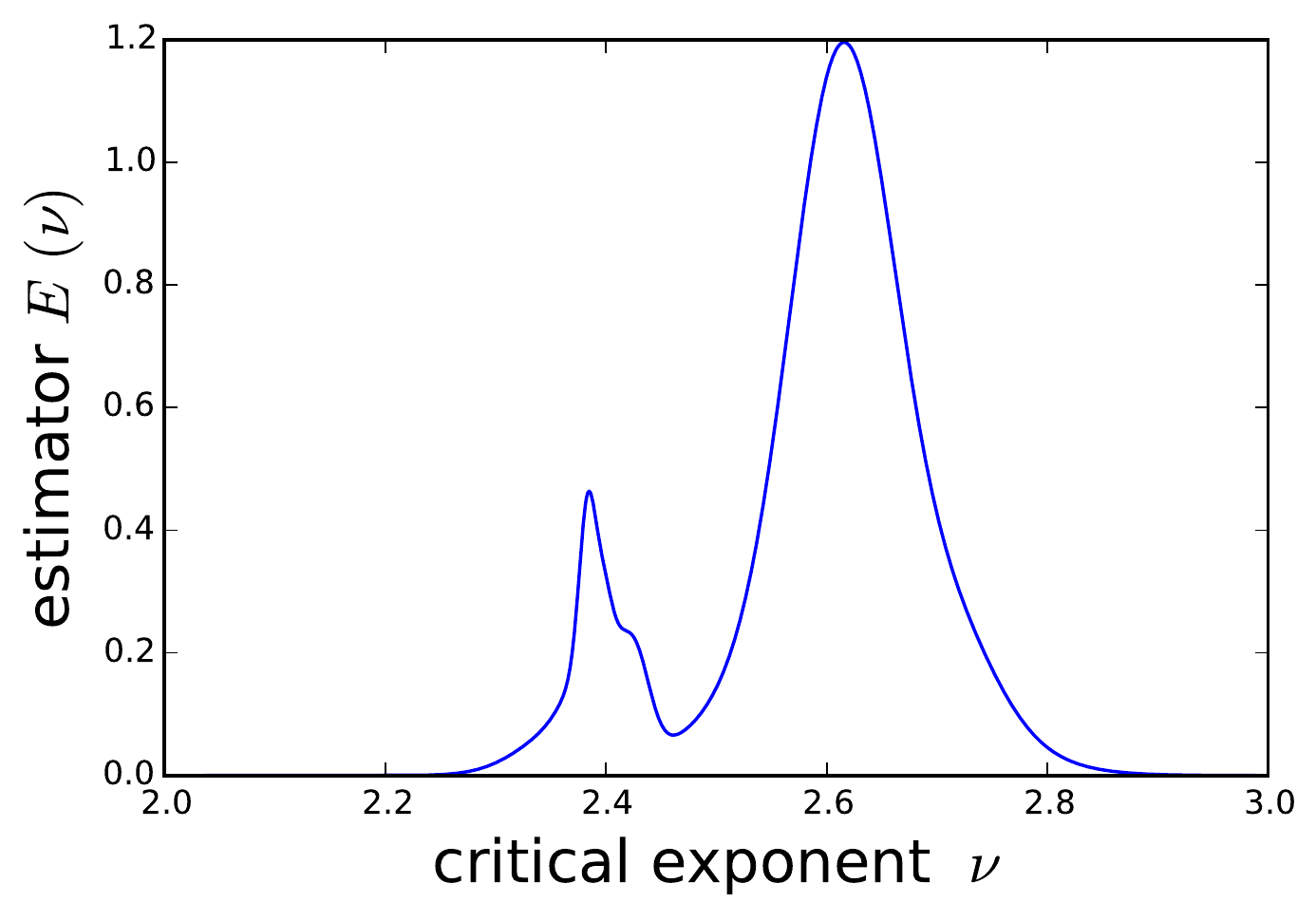}
  \includegraphics[width=\picwidth]{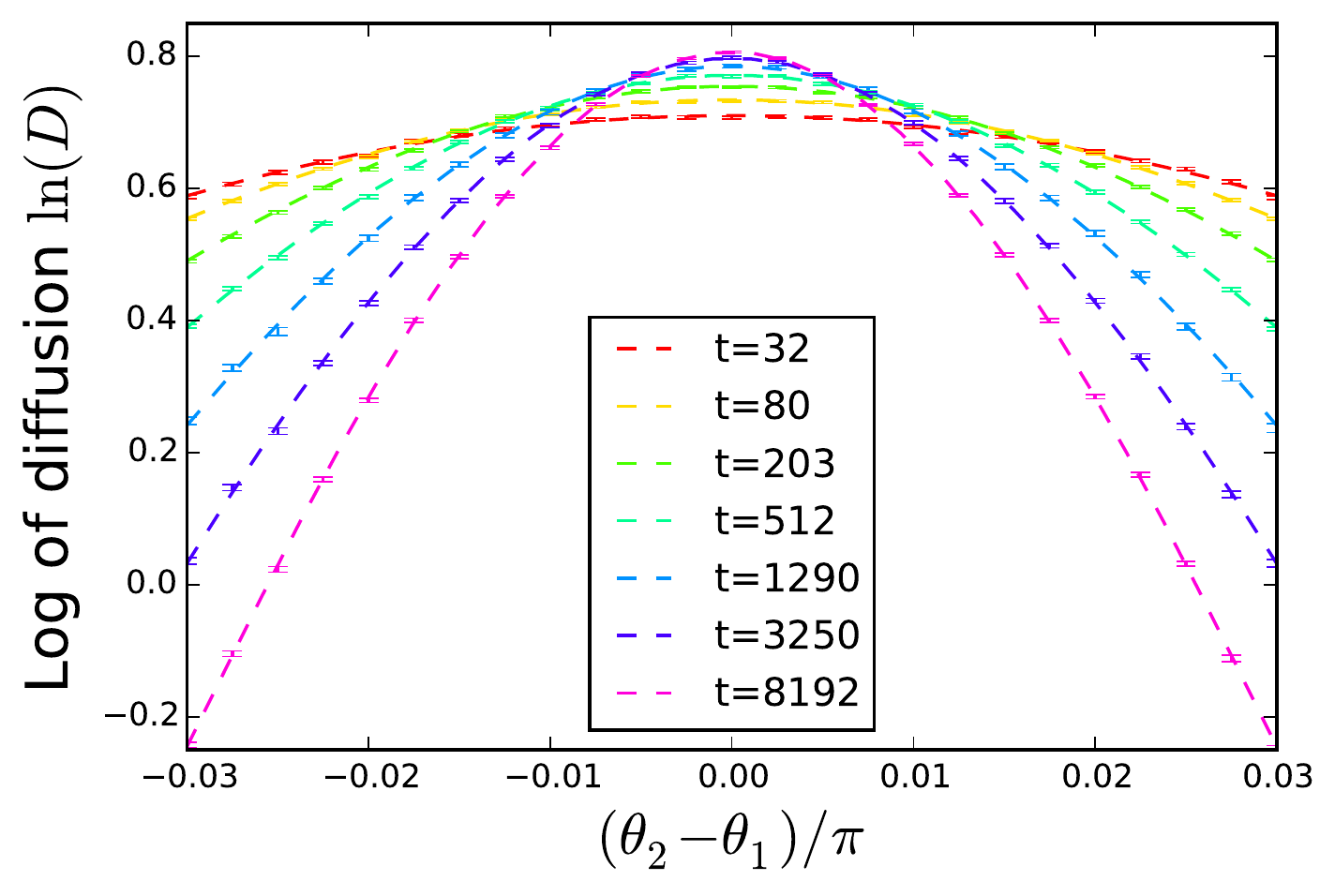}  
  \caption{(Color online) Top panel: estimator of critical exponent obtained. Rather than providing a histogram we have plotted the curve as a sum of 60 normalised Gaussians as given by eq.~\eqref{eq:def_of_nu_distribution} ($N_{max}=60$). In the bottom panel we show an example scaling fit to the diffusion coefficient data, demonstrating that the data for different $t$ and $\theta_2-\theta_1$ can be fitted to a single scaling function. Here $J=8, K=3$ and $L=1$ were chosen and $\nu=2.592$ at $\xi^2/ndf=0.94$ was obtained. The 68\% confidence interval was found to be $[2.557, 2.625]$.}
  \label{fig:distribution_of_nus}
\end{figure}

We computed the high accuracy data for the fitting procedure
by simulating the quantum walk on an $800 \times 800$ lattice for
varying number of timesteps over many disorder realizations.
A large number of disorder realizations was used for the runs at
shorter times (4001, 4001, 2001,1001, to obtain $D(t)$ at $t=32, 80,
203,512$, respectively), whereas due to self averaging, fewer disorder
realizations already provided enough accuracy for the runs at longer
times (200 realizations for $t =8192, 3250, 1290 $). The resulting
values of $D(t)$ were then fitted with the scaling Ansatz,
Eq.~\eqref{eq:scaling_ansatz_numeric}, using a Taylor series expansion
of the functions $F,G$, and $u$ to various orders.  Instead of
converging to a single solution, we obtained a good fit to the data
for different forms of the scaling functions, and also different
values of the exponents -- an example is shown in
Fig.~\ref{fig:distribution_of_nus}, lower panel.
To represent our estimate of the critical exponent $\nu$, we define an
estimator function $E(\nu)$, whose integral between any two values
$\nu_\text{min}$ and $\nu_\text{max}$ reflects our degree of
confidence that  
$\nu_\text{min} < \nu < \nu_\text{max}$. The construction of this function,
along with the details of the fitting procedure, are explained in
Appendix~\ref{sec:deta-scal-analys}.

As seen in Fig.~\ref{fig:distribution_of_nus}, our estimator of the
critical exponent $\nu$ is a bi-modal function, with a peak around
$\nu_m=2.616 $ (full width at half the maximum of $0.125$), and a
second peak at $ \nu_2=2.384$. The value corresponding to the larger
peak, $\nu_m =2.616$, is very close to the quantum Hall critical
exponent\cite{Slevin2009} of $2.593 \pm 0.003$. The smaller peak is
close to previous estimates of the exponent of the quantum Hall
transition\cite{Evers2008}, which are now attributed to bi-stability
of the fitting procedure, possibly related to finite-size
effects\cite{Slevin2012}.  To summarize, the transition which we
observe is compatible with the integer quantum Hall transition
universality class.

\section{Disorder in the rotation angles of the split-step walk}
\label{sec:disorder_angle}

We already introduced disorder to the rotation angles of the
split-step quantum walk, although with a small value of $\delta \theta=0.2\pi$, in the previous
sections. We now examine what happens to the quantum walk as this
disorder grows. We first consider a split-step quantum walk that is
localised by maximal phase disorder. As we turn on the angle disorder
$\delta \theta$, we will find that at special values of $\delta
\theta$, the walk delocalises. We then consider a split-step walk with
no phase disorder, only angle disorder. We find that, contrary to what
one might expect, angle disorder does not induce localisation.

\subsection{Competition of phase and angle disorder: 
Disorder-induced delocalisation}
\label{sec:another-way-drive}

We now consider what happens if we first localise a quantum walk by
phase disorder, as in section \ref{sec:local-due-onsite}, and then
increase the disorder in the rotation angles $\delta \theta = \delta
\theta_1 = \delta \theta_2$ to $\pi$.  At this maximal value, as well
as at $\delta \theta= \pi/2$, all inequivalent values of the rotation
angles are equally likely.
According to the network model pictured described in Sec.~\ref{sec:top_trans-mean_rot_angles}, we expect a percolating network of edge states and
thus expect delocalised
behaviour at these values of the rotation angle disorder.

\begin{figure}[tb]
  \centering \def \picwidth {\columnwidth}
  \includegraphics[width=\picwidth]{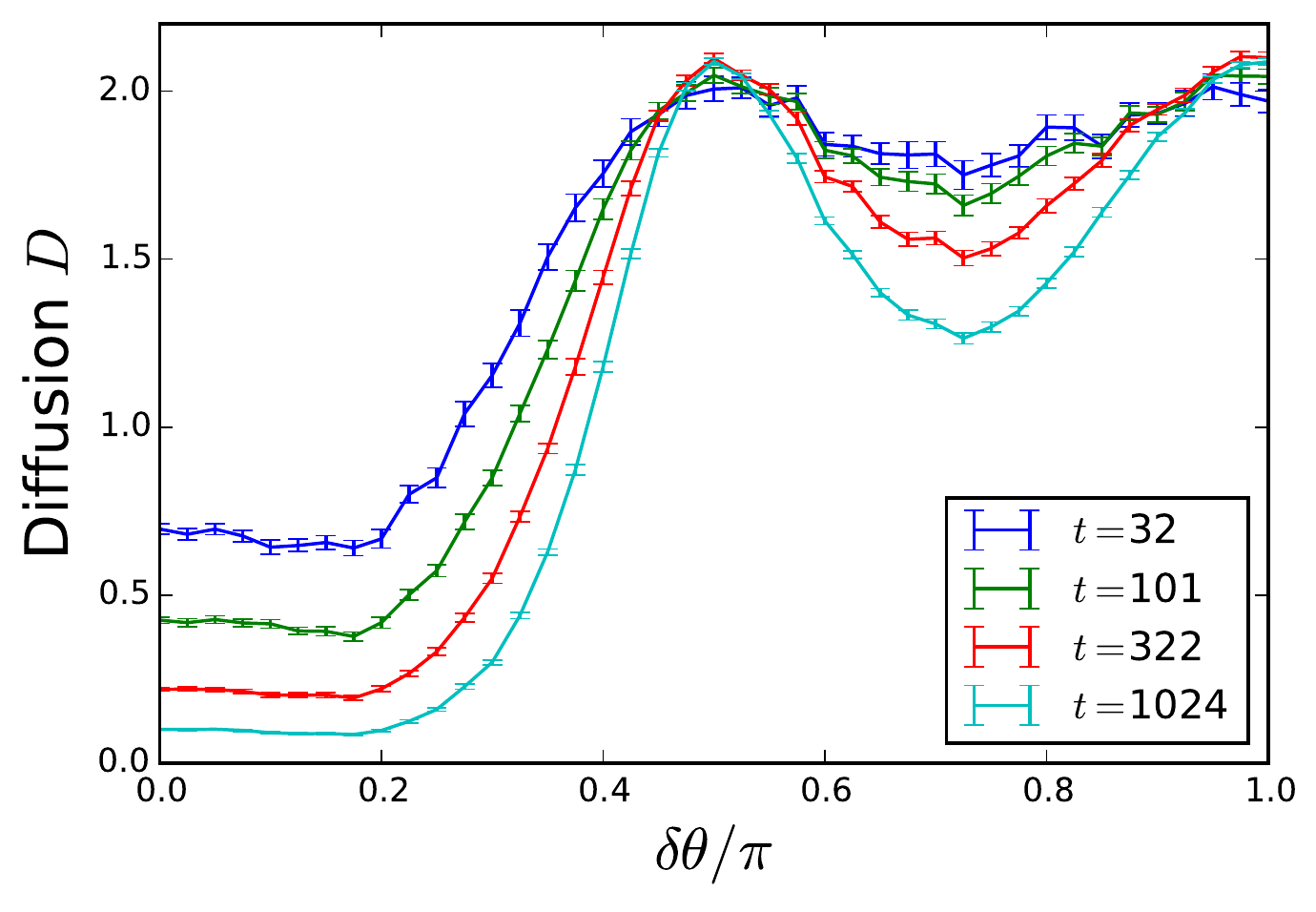}
  \caption{(Color online) Diffusion coefficient as a function of $ \delta\theta$ for $\delta\phi=2\pi$,  $\theta_1=0.35, \theta_2=0.15$, obtained by averaging over 100 disorder realisations on a 220x220 lattice.}
  \label{fig:diffn_coeff_whole_range}
\end{figure}

Our numerics clearly show the disorder-induced delocalisation, at both 
$\delta\theta=\pi/2$ and $\delta\theta=\pi$.  We
plot the time-dependent diffusion coefficient $D(t)$ in
Fig.~\ref{fig:diffn_coeff_whole_range}, as a function of $\delta\theta
$, at fixed mean values of the rotation angles, $\theta_1=0.35 \pi$,
$\theta_2=0.15 \pi$, and maximal phase disorder, $\delta \phi = 2 \pi$.
The diffusion coefficient decreases with time,
indicating localised dynamics, except near the points of maximal
disorder, $\delta\theta = \frac\pi2$ and $\pi$: there the system is
diffusive. 

We believe that the disorder-induced delocalisation we observe here
accompanies a topological phase transition, much like in the case of
Fig.~\ref{fig:diffn_coeff_fn_of_eta-whole_range}. Indeed, for $\delta
\theta < \pi/2$ the majority of sites have parameters corresponding to
topological invariants of $(+1,0)$, whereas for $\pi/2< \delta\theta <
\pi $, the majority topological invariant is $(-1,0)$.
It is thus plausible that at $\delta\theta=\pi/2$ a topological phase
transition occurs.
We performed a scaling analysis on this
transition, with now the control parameter being $\eta =
\delta\theta-\pi/2$. We obtained consistent results of $\nu=2.58\pm
0.05 $, in agreement with the mode of the distribution of $\nu_m=2.6 $
shown in Fig.~\ref{fig:distribution_of_nus}.  This confirms that the
exponent $\nu$ is universal: its value does not depend on the method
we use to drive the system across the transition.

\subsection{ Diffusive behaviour in the presence of only rotation angle disorder}
\label{sec:disorder_rotation}

\begin{figure}[tb]
  \centering \def \picwidth {\columnwidth}
  \includegraphics[width=\picwidth]{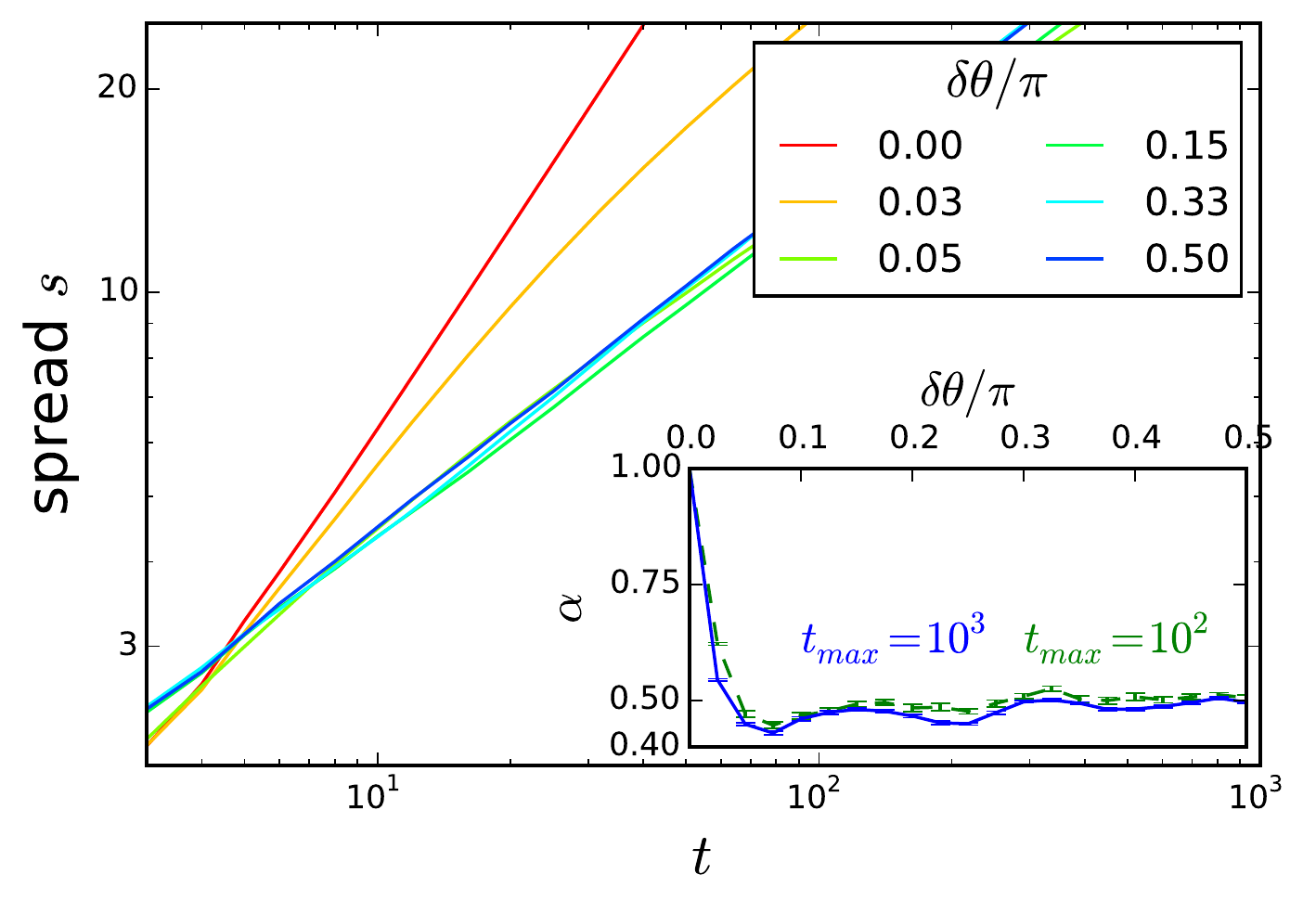}
  \caption{(Color online) Spread of the quantum walker averaged over 100 disorder
    realisations, for $\theta_1=0.35\pi, \theta_2=0.15\pi$ in the
    absence of phase disorder, as the angle disorder is increased from
    0 to its maximal value of $\pi/2$. Just like
    Fig.~\ref{fig:xover_deloc_loc_as_fn_of_phi} the inset shows the
    fitted power law exponent of $s(t)\propto t^\alpha$. For large
    $\delta\theta$ we obtain a time-independent $\alpha\approx
    \frac12$, unlike in the case where phase disorder was present.}
  \label{fig:xover_ballisic-diffn}
\end{figure}

Finally we investigate the spreading of the split step quantum walk in the presence of only rotation angle disorder. We fix the mean rotation angles to $\theta_1=0.35\pi$ and $ \theta_2=0.15\pi$. This choice of the mean rotation
angles places the system in an insulating phase with topological
invariants (+1,0), as shown by the blue four-sided star in
Fig.~\ref{fig:top_phase_diag}.

In Fig.~\ref{fig:xover_ballisic-diffn} we show the result of
increasing the rotation angle disorder from $\delta\theta=0 $ to
$\delta\theta=2\pi $. We observe the expected ballistic behaviour at
$\delta\theta=0 $, and already for rather small values of
$\delta\theta $ we see the crossover to the diffusive regime with
$s\propto t^{1/2} $. Unlike in the case with phase disorder, though,
we don't observe any signs of localisation here. 

Although we do not have a complete explanation for this absence of
localisation, we believe it is related to the particle-hole symmetry
of the system, that is not broken by rotation angle disorder.  The
time evolution operator has only real elements in position basis, and
thus, the effective Hamiltonian possesses particle-hole symmetry
represented by complex conjugation. In time-independent lattice
systems, the presence of this symmetry leads to non-universal
behaviour, and in some cases to diffusion instead of Anderson
localisation\cite{Evers2008}.

\section{Summary and conclusion}
\label{sec:summary-conculsion}

To summarise, we have found that the Hadamard walk is not localised by
phase disorder, while generic split-step quantum walks are. We gave an
intuitive physical explanation for this difference, namely, that the
Hadamard walk is a split-step walk tuned to a topological phase
transition. We corroborated this picture by numerically demonstrating
that this transition can be reached through angle disorder as well, at
precisely the value that this explanation predicts.  We determined the
critical exponent for the divergence of the localisation length for
both of these routes to criticality, and found $\nu=2.6$, which places
the split-step quantum walk with phase disorder in the universality
class of the quantum Hall effect.  We have also found that angle
disorder alone does not localise the split-step quantum walk, which
may be due to the fact that this disorder does not break the
particle-hole symmetry of the system.

A useful next step to strengthen our interpretation of the
localisation effects of disorder would be the calculation of the
topological invariant of the disordered split-step walk, the
quasienergy winding. Here, any of the existing approaches to the Chern
number in disordered systems can be of use.  One could extend the
definition of the quasienergy winding\cite{Rudner2012} using
noncommutative geometry\cite{prodan2011,Hastings2010}, or measure the winding
number of the scattering matrix\cite{fulga_2012b,Dahlhaus2011,Pasek2014}.

Another interesting question to pursue concerns at which point (and
whether) the disorder-driven delocalisation transition occurs for
non-uniform disorder distributions, e.g. gaussian disorder
distributions for $\theta_1$ and $\theta_2$ or a binary distribution
(with the two sets of $(\theta_1,\theta_2)$ having different
topological invariants).

Our interpretation of the localization phenomena relied on qualitative
similarity with disordered quantum (anomalous) Hall insulators: we
even obtained the same critical exponent. However, there are also ways
in which these two disordered systems differ from each other. In the
quantum anomalous Hall insulator study, a disorder-induced splitting
of the transition from Chern number $+1$ to $-1$ into two transitions
was observed \cite{Dahlhaus2011}. In the present paper, on the other
hand, no such splitting of the trasition from quasienergy winding $+1$
to $-1$ was found.  To better understand these differences, perhaps
the 4-step walk\cite{asboth_edge_2014a} can help, as it can realize
all possible combinations of trivial/nontrivial Chern number and
quasienergy winding.

\begin{acknowledgements}
We would like to acknowledge useful discussions with Mark Rudner,
Carlo Beenakker and Cosma Fulga. We also acknowledge the use of the
Leiden computing facilities. This work was funded by Nordita and also
supported by the Hungarian National Office for Research and
Technology under the contract ERC\_HU\_09 OPTOMECH, by the Hungarian
Academy of Sciences (Lend\"ulet Program, LP2011-016), and by by the
Hungarian Scientific Research Fund (OTKA) under Contract Nos. K83858
and NN109651.
\end{acknowledgements}

\appendix

\section{Details of the scaling analysis}
\label{sec:deta-scal-analys}

We determine the critical exponent $\nu$ by a scaling analysis as used
in Refs.~\onlinecite{Dahlhaus2011,Edge2012}, which will allow us to
classify to which universality class this localisation transition
belongs.  Instead of measuring the localisation length $\xi$ directly
from the numerics, we assume single parameter scaling, and determine
$\xi$ from the diffusion coefficient $D(t)$ of Eq.~\eqref{eq:2}.  The
scaling law for the logarithm of the diffusion coefficient in
dynamical localisation\cite{Lemarie2009-PRA} reads
\footnote{We choose $\ln D$ instead of $D$ to perform 
scaling on, since this makes the fitting simpler, 
as only a lower order expansion of the fitting 
function is required. This follows Ref.~\onlinecite{Dahlhaus2011}}
\begin{align}
  \label{eq:scaling_law_2d}
  \ln D(t) = \tilde F(\xi^{-2}t).
\end{align}
Here the scaling function $\tilde F(z)$ is some continuous,
differentiable function of its argument $z$.

We insert the power law diverging behaviour for $\xi$, given by
Eq.~\eqref{eq:def_nu}, into Eq.~\eqref{eq:scaling_law_2d}, rescale the
function $\tilde F$, and add finite-time
corrections\cite{Lemarie2009-PRA} to obtain
\begin{align}
  \label{eq:3}
  \ln D(t)&=F(t^{1/2\nu} u) + t^{- y}G(t^{1/2\nu} u); \\ 
    u&=\eta + {\cal O}(\eta ^2).
\end{align}
Here the function $F$ is related to $\tilde F$ as
\begin{align}
  \label{eq:5}
  F(z)= \tilde F(z^{2\nu}/A^2),
\end{align}
where $A$ is the constant from Eq.~\eqref{eq:def_nu}. 
The function
$G$ takes into account finite-time corrections, with
$ y$ denoting the first subleading exponent.  We expand the formulae
for $\ln D$ and $u$ of Eq.~\eqref{eq:3} in Taylor series,
\begin{align}
  \ln D &= \sum_{j=0}^{J} f_j (t^{1/2\nu} u)^{j} + 
\sum_{k=1}^{K} t^{-y} g_k (t^{1/2\nu} u)^{k -1} \nn\\
  u &=\eta + \sum_{l=3}^{L} u_l \eta^{l}. \nn
\end{align}
Since the function $\ln D$ must be even, $j$ may only take even
values. In contrast, $k$ and $l$ may only take odd values, though
$k=0$ is also allowed, $k=0$ corresponding to the absence of
finite-time corrections. 
We choose the order of the approximation by fixing $J, K, L
\in \mathbb{N}$, and then fit the Taylor coefficients $f_j$, $g_k$,
$u_l$, and the exponents $ y$ and $\nu$ to the $D(t)$ data. 
This allows us to obtain an estimate for the critical exponent $\nu$.

We fitted the numerically obtained data for $D(t)$
with 90 different functions, defined by different values of
$J\in\{2\dots10$\}, $K\in\{0,1,3\dots 9\}$, and $L\in\{1,3,5\}$. In a first
approach, we systematically increased the order of the approximation,
i.e., the values of $J, K $ and $L$, until we obtained a reasonable
goodness of fit.  Unfortunately, this did not yield a uniform
convergence, neither when the standard $\chi^2$-test was used (value
of $\chi^2$ per degree of freedom ($\chi^2/ndf $) of order 1), nor
when the more sophisticated goodness of fit measure\cite{Press2007},
$Q$, was used.
We thus resorted to an alternative approach, as explained below.

We represent our results for the critical exponent $\nu$, by use of an
estimator function $E(\nu)$, obtained by the following procedure.  Out
of the 90 different fitting functions ${\mathcal F}_i$, we reject
those which gave a value of $\chi^2$ outside of an acceptance range,
$0.5<\chi^2/ndf<2 $. For the remaining 60 functions ${\mathcal F}_i$,
with critical exponents $\nu_i$, we used the bootstrap method to
evaluate the goodness of fit $Q_i $, and the 68\% confidence interval
for the critical exponent: $\nu \in [\nu_i - \sigma_i^-, \nu_i +
  \sigma_i^+] $.
The estimator $E(\nu)$ is then defined as
\begin{align}
  E(\nu) &= \frac1{N_{max}} \sum_{i=1}^{N_{max}} \sqrt{2\pi\sigma_i} \exp
  \left(
    -\frac{(\nu-\nu_i)^2}{2\sigma_i^2}
  \right),
  \label{eq:def_of_nu_distribution}
\end{align}
with $N_{max}=60$, and $\sigma_i = (\sigma_i^+ + \sigma_i^-)/2$. This
is the probability density of the critical exponent, if we deem all
acceptable outcomes of our fitting procedure equally likely.


%

\end{document}